\documentclass[sigplan, 10pt]{acmart}
\acmConference[PPoPP '25]{30th ACM SIGPLAN Annual Symposium on Principles and Practice of Parallel Programming 2025}{March 01--05, 2025}{Las Vegas, NV, USA}
\acmYear{2025}
\acmISBN{} 
\acmDOI{} 
\startPage{1}

\setcopyright{none}

\usepackage{balance, subcaption, multicol, listings, todonotes}
\usepackage[ruled, vlined]{algorithm2e}

\usepackage{xcolor}

\definecolor{codegreen}{rgb}{0,0.6,0}
\definecolor{codegray}{rgb}{0.5,0.5,0.5}
\definecolor{codepurple}{rgb}{0.58,0,0.82}
\lstdefinestyle{mystyle}{
    commentstyle=\color{codegreen},
    keywordstyle=\color{magenta},
    numberstyle=\tiny\color{codegray},
    stringstyle=\color{codepurple},
    basicstyle=\ttfamily\scriptsize,
    breakatwhitespace=false,         
    breaklines=true,                 
    captionpos=b,                    
    keepspaces=true,                 
    numbers=left,                    
    numbersep=5pt,                  
    showspaces=false,                
    showstringspaces=false,
    showtabs=false,                  
    tabsize=2
}
\begin{document}

\title{Skip Hash: A Fast Ordered Map Via Software Transactional Memory}

\author{Matthew Rodriguez}
\email{mrodriguez@commonwealthu.edu}
\affiliation{%
  \institution{Commonwealth University of Pennsylvania}
  \city{Bloomsburg}
  \state{Pennsylvania}
  \country{USA}
}

\author{Vitaly Aksenov}
\email{aksenov.vitaly@gmail.com}
\affiliation{%
  \institution{City, University of London}
  \city{London}
  \country{UK}
}

\author{Michael Spear}
\email{spear@lehigh.edu}
\affiliation{\institution{Lehigh University}\city{Bethlehem}\state{Pennsylvania}\country{USA}}

\begin{abstract}

Scalable ordered maps must ensure that range queries, which operate over many consecutive keys, provide intuitive semantics (e.g., linearizability) without degrading the performance of concurrent insertions and removals.
These goals are difficult to achieve simultaneously when concurrent data structures are built using only locks and compare-and-swap objects.
However, recent innovations in software transactional memory (STM) allow programmers to assume that multi-word atomic operations can be fast and simple.

This paper introduces the skip hash, a new ordered map designed around that assumption.
It combines a skip list and a hash map behind a single abstraction, resulting in $O(1)$ overheads for most operations.
The skip hash makes use of a novel range query manager---again leveraging STM---to achieve fast, linearizable range queries that do not inhibit scalability.
In performance evaluation, we show that the skip hash outperforms the state of the art in almost all cases.
This places the skip hash in the uncommon position of being both exceedingly fast and exceedingly simple.
\end{abstract}



\maketitle

\section{Introduction}
\label{sec-intro}

Ordered maps are among the most fundamental data structures for modern data-intensive applications.
These data structures implement \emph{elemental} operations for creating, updating, clearing, and querying associations between keys and values.
They also support \emph{point query} operations such as \texttt{floor}, \texttt{ceil}, \texttt{pred}, and \texttt{succ}, which find the closest key in the map that is respectively $\leq$, $\geq$, $<$, or $>$ a given key.
Finally, they support a \emph{range query} operation, which gathers all key/value pairs whose keys fall within some range $[l,r]$.

Applications must employ many threads to process large amounts of data quickly, and so they must use ordered maps that scale well.
%
%
However, it is challenging to implement efficient range queries that provide a reasonable correctness property (e.g., linearizability~\cite{herlihy-toplas-1990}).
%
%
%
%
%
Consider an execution history where one thread executes \texttt{range(l, r)} while other threads repeatedly try to insert and remove keys between \texttt{l} and \texttt{r}.
If the range query does not block elemental operations, then how can it ensure that its result corresponds to the map's state at some point in time?

Prior work has largely coalesced around two ideas.
First, optimistic synchronization techniques~\cite{herlihy-book-2020} are employed so that elemental operations do not induce artificial contention when reading data structure nodes.
Second, some manner of versioning is attached to map entries, so that range queries can operate over a ``frozen'' version of the data structure while concurrent elemental operations update a ``fresh'' copy.

The key insight in this paper is that modern software transactional memory (STM) systems have become surprisingly efficient~\cite{brown-ppopp-2022, timnat-europar-2015, guerraoui-ppopp-2016, sheng-pact-2023, felber2010time}.
This is important, because STM simplifies two common synchronization behaviors: (1) modifying several locations as a single, indivisible operation and (2) performing an action only if some variable remains unchanged throughout that action's duration.

Rather than ask how STM can improve existing concurrent ordered maps, in this paper we design a map from scratch using STM as its primary synchronization mechanism.
Our first observation is that STM lets us compose a hash map with a \emph{doubly linked} skip list.
We call this data structure a ``skip hash.''
The skip hash scales well, has $O(1)$ complexity for most elemental operations, and is easy to verify.
However, it is prone to starvation for long-running range queries.
We remedy this by introducing a new mechanism for coordinating range queries and elemental operations.
This mechanism adds $O(1)$ overhead to elemental operations (typically only a single read).
However, it has high contention when range queries are small or many removals execute concurrently with a range query.
Finally, we employ a fast-path/slow-path strategy~\cite{kogan-ppopp-2012}, so that the range query coordinator is only used as a fallback, alleviating this contention.

On standard microbenchmarks, the skip hash outperforms the current state of the art on almost all configurations, often by a large margin.
In addition, since the skip hash uses STM, its implementation is free from complex and error-prone synchronization.
Furthermore, STM makes it trivial for the skip hash to support complex key and value types beyond those that fit in a single memory word.
We conclude that the skip hash and the STM-first design methodology from which it emerged provide a compelling baseline for future concurrent data structure research and practice.


\section{Related Work}
\label{sec-related-work}

\subsection{Range Queries}

One of the main advantages of ordered maps is their ability to efficiently support range queries. 
There is a significant body of work on how to implement these queries in a concurrent setting~\cite{brown2012range, arbel-ppopp-2018, wei2021constant, miller2024brief, nelson2022bundling, zhong2021remix, basin2017kiwi}, including a recent focus on aggregate range queries~\cite{kokorin2024wait, fatourou2024lock, sela2024concurrent}. 
Early works in this area often relied on restarting a range query when a concurrent update occurred. 
One particularly relevant study~\cite{avni-podc-2013} used STM for this purpose. 
However, at the time, STM implementations were neither highly efficient nor expressive, which limited scalability, especially in experiments with a high update rate.

Arbel and Brown utilized an epoch-based reclamation (EBR) scheme~\cite{arbel-ppopp-2018} in their range queries. 
The key idea is to assign a timestamp to each range query, enabling it to determine whether an element existed before it began. 
The main challenge arises with removal operations, as a needed element may be missed by the traversal if it is concurrently removed. 
Their technique leveraged EBR, in which removed elements are temporarily stored in a ``limbo set,'' to detect overlooked elements after the range query completed.
This concept is, in a sense, the inverse of our design, where removal of elements is \emph{delegated to} range queries, thereby avoiding a search through the limbo set.

The general approach in most prior work, such as vCAS~\cite{wei2021constant} and Bundling~\cite{nelson2022bundling}, is based on Multi-Version Concurrency Control (MVCC). 
In these methods, updates to a data structure do not overwrite the previous versions of updated elements, links, etc.
Since space overheads can grow significantly, custom garbage-collection-like mechanisms are then used to reclaim versions when they are no longer useful.
One work~\cite{miller2024brief} that follows this approach is particularly interesting because it uses hash-tries as the foundation instead of the usual ordered sets. 
However, the issue with hash tries and similar structures~\cite{oshman2013skiptrie, areias2018extending, prokopec-ppopp-2012, prokopec2018cache, miller2024brief} is that they can only support range queries if the key itself is used as the hash. 
This is not always feasible and can result in an unbalanced structure. 
Among these works,~\cite{prokopec2018cache} employs an idea similar to ours, using a hash map as a cache, but in a different context.

\subsection{Modern STM Systems}
TM was originally proposed as a technique for simplifying the creation of concurrent data structures~\cite{herlihy-isca-1993, shavit-podc-1995}.
While there was considerable effort to expand TM's scope so that it could serve as the foundation of a full-fledged programming model~\cite{ni-oopsla-2008}, there have always been arguments that it should be thought of as merely an ``implementation technique''~\cite{boehm-hotpar-2009}.
This has two main benefits.
First, when programmers can constrain the scope of the TM, they can guarantee that transactional data is never accessed outside of transactions.
The resulting transactional semantics~\cite{abadi-popl-2008} avoid ``privatization'' overheads~\cite{menon-spaa-2008}.
Second, if programmers are allowed to know about the TM's internal implementation, they can employ many low-level optimizations~\cite{dragojevic-eurosys-2012} to reduce latency.

In recent years, several STM systems have explored this design space, including TinySTM~\cite{felber2010time}, MCMS~\cite{timnat-europar-2015}, Lock-Free Locks~\cite{ben2022lock}, PathCAS~\cite{brown-ppopp-2022}, exoTM~\cite{sheng-pact-2023}, and the slow clock STM~\cite{ramelhete-ppopp-2024}.
While these systems differ with regard to the offered progress guarantees, they coalesce around the following design principles:
\begin{itemize}
\item STM algorithms based on ownership records (orecs)~\cite{dice-disc-2006, saha-ppopp-2006} scale best for map data structures.
\item The use of a global clock within the STM algorithm~\cite{dice-disc-2006} need not be a significant bottleneck~\cite{lev-spaa-2009, ruan-taco-2013, ramelhete-ppopp-2024}.
\item Acquiring orecs upon first write access~\cite{saha-ppopp-2006, dice-cgo-2007, felber-ppopp-2008} and leveraging undo logging results in the lowest latency.
\item Static read-only transactions can be optimized to negligible overhead~\cite{dice-disc-2006}.
\item Orecs should be co-located with the objects they protect, not kept in a separate table~\cite{brown-ppopp-2022, timnat-europar-2015, sheng-pact-2023, ramalhete-ppopp-2024}.
\item Programmers should optimize instrumentation for \texttt{const} fields and use specialized API calls to avoid redundant logging~\cite{riegel-spaa-2008, dragojevic-spaa-2009, harris-pldi-2006}.
\item STM instrumentation should be in-lined into the program~\cite{bendavid-ppopp-2022, brown-ppopp-2022, ramelhete-ppopp-2024}.
\end{itemize}
Finally, these works have collectively shown that STM can be fast, even without hardware acceleration.

\section{A Fast Ordered Map}
\label{sec-elemental-design}

The best-performing lock-free map implementations use compare-and-swap objects~\cite{fraser-phd-2003, srivastava2022elimination, brown2020non, natarajan2014fast}.
To simplify linearization arguments, these maps are singly linked.
That is, nodes in tree-based maps tend not to have parent pointers, and nodes in skip lists tend not to have predecessor pointers.
However, STM makes it straightforward to implement double-linking and to access multiple data structures in a single, indivisible operation.
Given these benefits of STM, Figure~\ref{lst-skiphash} defines the skip hash as the composition of a doubly linked skip list that maps keys to values, and a closed-addressing hash table that maps keys to skip list nodes.

\begin{figure}
\vspace{-4mm}
\begin{lstlisting}[escapechar=`, firstnumber=1]
type sl_node<K, V>     # A node in the skip list `\label{lst-sl-node-def}`
  const key: K         # The (immutable) key
  val: V               # The associated value
  const height: u8     # Height of "neighbors" (min 1)
  # Array of predecessor/successor links at each level
  neighbors: Array<(sl_node<K, V>, sl_node<K, V>)>

type skiplist<K, V>
  head: sl_node<K, V> # Sentinel nodes initially stitched`\label{lst-sl-head}`
  tail: sl_node<K, V> #   together at all levels `\label{lst-sl-tail}`

type skip_hash<K, V>
  map: hashmap<K, sl_node<K, V>>
  sl: skiplist<K, V>

func lookup(k: K) -> Option<V>: `\label{lst-lookup}`
  atomic:
    # If the key is present, the map routes to the node
    let n = map.get(k)
    if (!n) return None else return n.val

func remove(k: K) -> bool:
  atomic:
    # If the key is not present, stop at map lookup
    let n = map.get(k) `\label{lst-norange-remove-hashmap-get}`
    if (!n) return false `\label{lst-norange-remove-absent}`
    # Otherwise remove from the map, and leverage 
    #   double-linking to avoid skip list traversal
    map.remove(k)
    n.unstitch()
    return true

func insert(k: K, v: V) -> bool: `\label{lst-norange-insert}`
  atomic:
    # O(1) if the key is already present
    if (map.get(k)) return false
    # O(log n) with optimized skip list insert,
    #   because k is necessarily absent
    let new_node = sl.insert_optimized(k, v) `\label{lst-norange-insert-optimization}`
    # Finally, update the map to reference the new node
    map.insert(k, new_node)
    return true

func ceil(k: K) -> K: # Find smallest key >= k `\label{lst-ceil}`
  atomic:
    if (map.get(k)) return k # O(1) if key present
    return sl.succ(k).key    # O(log n) otherwise

func succ(k: K) -> K: # Find smallest key > k `\label{lst-succ}`
  atomic:
    let node = map.get(k)
    if (node) return node.neighbors[0].second.key
    return sl.succ(k).key
\end{lstlisting}
\vspace{-5mm}
\caption{Transactional Composition of an Unordered Map and Skip List}
\label{lst-skiphash}
\end{figure}

The skip list is implemented as a sequence of $n$ objects of type \texttt{sl\_node} (line \ref{lst-sl-node-def}).
Each contains a key/value pair $(k, v)$, a height $h$, and a ``tower'' consisting of $h$ pairs of pointers.
Upon each node's insertion, $h$ is generated using a geometric distribution with $p=1/2$ in the range $[1,m]$ where $m \geq \lg{n}$.
The skip list has $m$ levels, where each level $l$ represents a doubly linked list whose traversal ``skips'' all nodes where $h \leq l$.
Sentinel head and tail nodes (lines~\ref{lst-sl-head}--\ref{lst-sl-tail}) of height $m$, with keys $\bot$ and $\top$ respectively, bookend the skip list.
%

The skip hash is designed with the invariant that the set of keys present in the skip list and hash map is identical at all times.
The use of this property asymptotically accelerates several skip list operations.
Furthermore, we assume without loss of generality that all hash map operations  are $O(1)$.  
Therefore, we observe that $\texttt{lookup(k)}$ (line \ref{lst-lookup}) is always $O(1)$, consisting of a map lookup and then at most one additional read.
When $k$ is absent, $\texttt{remove(k)}$ is also $O(1)$, returning immediately after the failed map lookup (line \ref{lst-norange-remove-absent}). 
The same logic explains why $\texttt{insert(k)}$ (line \ref{lst-norange-insert}) and point queries are all $O(1)$ when $k$ is present. 
This can be seen in \texttt{ceil(k)} (line~\ref{lst-ceil}) and \texttt{succ(k)} (line~\ref{lst-succ}); \texttt{floor(k)} and \texttt{pred(k)} can be implemented similarly.

When $k$ is present, $\texttt{remove(k)}$ avoids an $O(\log{n})$ skip list traversal by using the map to find the node with key $k$ (line \ref{lst-norange-remove-hashmap-get}).
Since the skip list is doubly linked, the predecessor and successor at each level can be located in $O(1)$ time, so unstitching any one level has constant overhead.
The expected height of a randomly selected node is $1/p=2$, so the average complexity is $O(1)$, with a worst case of $O(\log{n})$.  

Thus, the only operations that cannot avoid an $O(\log{n})$ skip list traversal are $\texttt{insert(k)}$ and the four point queries, and only when when $k$ is absent.  
Point queries then perform $O(1)$ additional reads, and $\texttt{insert()}$ performs up to $O(\log{n})$  writes.
However, the expected height is $O(1)$, like with $\texttt{remove()}$; and since the key is guaranteed to be absent, the skip list insertion logic can be optimized (line \ref{lst-norange-insert-optimization}).
Note that the presence of predecessor pointers doubles the number of writes compared to a singly linked skip list.

To summarize, STM lets us implement the map interface by composing two data structures.  
The resulting data structure avoids skip list traversal in most cases, and thus the majority of its operations only perform a constant number of reads and (expected) writes.

When using a modern STM as described in Section~\ref{sec-related-work}, the skip hash's ability to scale hinges on the probability of concurrent operations accessing the same datum at the same time, with at least one performing a write.
Because transactions in a skip hash must validate after performing a write, they have a larger contention window than a skip list implemented with lock-free or optimistic locking techniques.
There is a surprising caveat, however: $\texttt{remove()}$ operations do not read any skip list node that they do not also write.  Therefore, $\texttt{remove()}$ is not vulnerable to the sorts of artificial aborts that TM is known to induce.
The only situation where unnecessary conflicts can arise, then, is between pairs of insertions.  Our evaluation shows that this is not a significant concern.

\section{Supporting Range Queries}
\label{sec-range-design}



Given our use of STM, the simplest implementation of a linearizable range query is to execute the entire operation as a single transaction. 
This approach avoids introducing any global synchronization metadata, so range queries will only conflict with concurrent overlapping insertions and removals. 
Such conflicts should be uncommon in low-skew workloads where the data structure is highly populated, updates are rare, or range queries are short. 
In workloads where conflicts are more common, however, STM-based range queries may require many attempts to succeed. 
This can significantly degrade performance or even lead to the sort of starvation that cannot be solved through traditional contention management~\cite{bobba-isca-2007}. 

Therefore, we introduce a new object in this section: the range query coordinator (RQC).
The RQC assigns version numbers to range queries and skip list nodes.
These versions allow a range query to ignore nodes inserted after or removed before it began.
We can then execute a range query as a sequence of transactions, each of which accesses a small number of consecutive skip list nodes.

While this approach does not alter the implementation of \texttt{lookup()} and only trivially changes \texttt{insert()}, it creates a new burden for \texttt{remove()}:
While a node can be ``logically removed'' through the use of version numbers, it cannot be unstitched if a concurrent range query needs it.
Therefore, we introduce the concept of ``safe nodes'', which cannot be unstitched during an in-flight range query and which can serve as boundary points for a range query's transactions.
We also introduce a deferral mechanism through which \texttt{remove()} delegates unstitching to an in-flight range query.
While delegation only adds $O(1)$ overhead, it can increase contention.
We reduce contention via a fast-path strategy: 
Every range query first tries to complete along the fast path---finishing the entire query in a single transaction---before falling back to the slow path.
On the slow path, it uses one transaction to acquire a unique version number, a sequence of transactions to access the pairs in the map, and then a finalizing transaction.
By keeping transactions short, the range query is thus able to avoid contention and make progress.

\subsection{An Abstract Coordinator}

The foundation of our slow path is the RQC.
Abstractly, it provides four methods, summarized below:

\begin{itemize}
\item \texttt{on\_range()} -- Registers a new slow-path range query and assigns a unique version number to it.
\item \texttt{on\_update()} -- Reports the most recent range's version number to the calling  \texttt{insert()} or \texttt{remove()}.
\item \texttt{after\_remove(sl\_node)} -- Unstitches and reclaims a logically deleted skip list node immediately if safe; otherwise, schedules it for deferred removal.
\item \texttt{after\_range(ver)} -- Marks the range query with version number \texttt{ver} complete. May also unstitch and reclaim nodes whose removal was deferred by concurrent calls to \texttt{after}\-\texttt{\_remove()}.
\end{itemize}

The RQC has two obligations. First, it produces monotonically increasing version numbers which order slow-path range queries with respect to successful insertions and removals. 
Second, it determines when it is safe to unstitch and reclaim logically deleted nodes.

\subsection{Logical Deletion}
We augment the \texttt{sl\_node} type with two new fields. 
\texttt{n.i\_time} is an immutable value that records the version number of last range query that began before $n$'s insertion.
\texttt{n.r\_time} is initially $\texttt{None}$, indicating $n$ is logically present. 
To \emph{logically delete} $n$, this field is set to the most recent range query's version.
$n$ may remain ``physically'' linked into the skip list for some time afterwards to allow it to be processed by that range query and/or its predecessors.
For any other purpose, $n$ is considered semantically absent from the data structure.

While nodes may remain in the skip list for some time after their logical deletion, they are removed from the hash map immediately. 
This provides a powerful invariant: The hash map always reflects the current logical state of the data structure. 
This means no changes are needed to the \texttt{lookup()} operation.
When point queries do not find their operand in the map and therefore must perform a lookup in the skip list, they require a single-line edit to check \texttt{r\_time} and ensure they do not return a logically deleted node.
The $\texttt{insert()}$ and $\texttt{remove()}$ operations require slightly more invasive changes, which are shown in Figure~$\ref{lst-rqc-elementals}$.

\begin{figure}
\vspace{-4mm}
\begin{lstlisting}[escapechar=`, firstnumber=1]
func remove(k: K) -> bool:
  atomic:
    let n = map.get(k) `\label{lst-augmented-remove-hashmap-get}`
    if (!n) return false # Failed; key absent `\label{lst-augmented-remove-absent}`
    map.remove(k) `\label{lst-augmented-remove-hashmap-remove}`
    n.r_time = rqc.on_update() # Logically delete `\label{lst-set-rtime}`
  rqc.after_remove(n) `\label{lst-after-remove}`
  return true

func insert(k: K, v: V) -> bool:
  atomic:
    let n = map.get(k)
    if (n) return false # Failed; key already present
    let i_time = rqc.on_update() `\label{lst-set-itime}`
    # The key may be present in the skip list, 
    #   but only in a logically deleted node
    let new_node = sl.insert_after_logical_deletes(k, v)
    new_node.i_time = rqc.on_update()
    map.insert(k, rqc)
    return true
\end{lstlisting}
\vspace{-5mm}
\caption{Elemental operations of a skip hash augmented with a range query coordinator.}
\label{lst-rqc-elementals}
\end{figure}

Leveraging the invariant described above, \texttt{remove(k)} begins by querying the map for a node with key $k$ (line \ref{lst-augmented-remove-hashmap-get}). 
If $k$ is absent from the map, the operation completes and returns \texttt{false} (line \ref{lst-augmented-remove-absent}). 
Otherwise, the operation removes the node from the map to maintain the invariant (line \ref{lst-augmented-remove-hashmap-remove}) and logically deletes the node by setting \texttt{r\_time} to the current version number, as reported by \texttt{on\_update()} (line \ref{lst-set-rtime}). 
At this point, subsequent elemental operations will not find the key, and concurrent range queries will use the node's version number to decide whether to process it.
All that remains is to unstitch and reclaim the node.
This is accomplished via $\texttt{after\_remove(n)}$, which may choose to delegate clean-up to an ongoing range query (line \ref{lst-after-remove}).

Relative to Figure~\ref{lst-skiphash}, \texttt{insert(k)} changes in two ways.
When the hash map indicates that $k$ is absent, the operation uses \texttt{on\_update()} to initialize the new node's \texttt{i\_time} (line \ref{lst-set-itime}).
Additionally, the logic used to insert that node into the skip list must be modified slightly, because one or more logically deleted nodes with key $k$ may still be present in the skip list.
So, instead of completing as a failed insertion upon finding the key present in the skip list, it will instead insert the new node \emph{after} any logically deleted nodes with key $k$.

\subsection{Safe Nodes}
\label{subsec-safe-nodes}




In Figure~\ref{lst-rqc-elementals}, \texttt{insert()} and \texttt{remove()} can modify the skip list at all times.
While TM handles any conflicts that occur during a transaction, slow-path range queries must tolerate skip list changes when they are \emph{between} transactions. 
Suppose an ongoing slow-path range query $R$ commits a transaction partway through its range and stops on some node $n$. 
If a concurrent \texttt{remove()} operation deletes $n$, this could result in $R$ accessing freed memory or other erroneous behavior. 
To prevent this, we ensure that slow-path range queries only stop on \emph{safe nodes}---nodes guaranteed not to be unstitched or reclaimed during their execution.

For a query $R$ on range $(l, r)$ with version number $ver$ to be correct, it must process every node in its \emph{processing set}, defined as every node $n$ with the following properties:
\begin{enumerate}
\item $l \leq \texttt{n.key} \leq r$ (i.e., \texttt{n}'s key is in $R$'s range)
\item $\texttt{n.i\_time} < ver$ (i.e., \texttt{n} was inserted before $R$ began)
\item $\texttt{n.r\_time} = \texttt{None} \lor \texttt{n.r\_time} \geq ver$ (i.e., \texttt{n} was not deleted before $R$ began)
\end{enumerate}

For $R$ to linearize, $n$ must remain in the skip list until $R$ has progressed past it. 
We adopt a stricter condition that requires less inter-thread communication: $n$ cannot be removed until $R$ completes.
Thus, every node in $R$'s processing set is also a safe node where $R$ can stop between transactions. 
Additionally, any node $n$ which violates condition (1) but satisfies conditions (2)\footnote{The RQC may immediately unstitch nodes inserted after the most recent range query, so a node that violates condition (2) is not necessarily safe.} and (3) is also considered safe, despite being outside of the processing set.
Lastly, as the head and tail sentinel nodes are never removed from the skip list, they are also considered safe.

For any removed node $n$, if $n$ is needed by an ongoing slow-path range query $R$, the RQC defers $n$'s removal until $R$'s invocation of \texttt{after\_range()} at the earliest. 
In doing so, the RQC ensures that $n$ is safe for $R$.
This is why the RQC is responsible for unstitching and reclaiming nodes.

As soon as a range query $R$ commits a transaction in which it called $\texttt{on\_range()}$, the set of safe nodes for $R$ is fixed---no nodes can enter or leave it. 
Any newly inserted node will have $\texttt{i\_time} > ver$, violating condition (2).
Any removed safe node will have its $\texttt{r\_time}$ change from $\texttt{None}$ to a value $\geq ver$, preserving condition (3).
Thus a range query does not need to access every node in its processing set in a single transaction.  
Instead, it can split its operation into several transactions, each beginning and ending on a safe node.

\subsection{Implementing the Range() Operation}

\begin{figure}
\vspace{-4mm}
\begin{lstlisting}[escapechar=`, firstnumber=1]
# Process all nodes with a key in the range [l, r]
func range(l: K, r: K) -> Set<(K,V)>: `\label{lst-range-method}`
  # Try the fast path a few times
  for i in 0 .. FAST_PATH_TRIES: `\label{lst-fast-path-loop}`
    let set = range_fast(l, r)
    if (set != None) return set
  atomic:  # Fall back to slow path `\label{lst-slow-path}`
    let start = ceil(l)
    let ver = rqc.on_range()
  let set = range_slow(start, r, ver)
  rqc.after_range(ver) `\label{lst-slow-after-range}`
  return set

# Try to complete a fast-path range op
func range_fast(l: K, r: K) -> Option<Set<(K,V)>>: `\label{lst-range-fast-method}`
  atomic(try_once): # does not retry on conflict `\label{lst-try-once}`
    let set = {}
    let n = sl.ceil(l) `\label{lst-fast-path-ceil}`
    while n.key <= r:
      if (n.r_time == None) set.add((n.key, n.val)) `\label{lst-fast-add}`
      let n = n.neighbors[0].second
    return set
  return None
    
# Process nodes in range for a slow-path range query
func range_slow(n: sl_node, r: K, v: u64) -> Set<(K,V)>: `\label{lst-range-slow-method}`
  let set = {}
  # this atomic block won't undo changes to n/set
  atomic(no_local_undo):`\label{lst-no-local-undo}`
    while n.key <= r: `\label{lst-range-process-loop-start}`
      let next = next_safe(n, v)
      set.add((n.key, n.val))
      n = next `\label{lst-range-process-loop-end}`
  return set

# Find the next safe node after n for RQC value ver
func next_safe(n: sl_node, ver: u64) -> sl_node<K,V>: `\label{lst-next-safe-method}`
  let n = n.neighbors[0].second
  while !is_safe(n, ver) n = n.neighbors[0].second
  return n

# Determine if the given node is safe
func is_safe(n: sl_node, ver: u64) -> bool: `\label{lst-is-safe-method}`
  if (n == sl.head || n == sl.tail) return true
  if (n.i_time >= ver) return false
  return (n.r_time == None || n.r_time >= ver)

\end{lstlisting}
\vspace{-5mm}
\caption{Range query implementation}
\label{lst-range}
\end{figure}

Figure~\ref{lst-range} presents the full \texttt{range()} algorithm.
\texttt{range(l,r)} takes two keys representing the bounds of the range (line \ref{lst-range-method}).
First, the algorithm tries the fast path a few times (line \ref{lst-fast-path-loop}).
In our implementation, we set $\texttt{FAST\_PATH}\-\texttt{\_TRIES}$ to 3.
If that threshold is exceeded, it falls back to the slow path (line \ref{lst-slow-path}).

\subsubsection{Fast Path}
\texttt{range\_fast()} (line \ref{lst-range-fast-method}) attempts to execute the range query as a single transaction.
It does not need to worry about safe nodes and is not assigned a version number, so its logic is relatively simple.
We use \texttt{atomic(try}\-\texttt{\_once)} to indicate that the transaction should not retry if it aborts (line \ref{lst-try-once}).
This lets the caller fall back to the slow path after \texttt{FAST\_PATH\-\_TRIES} attempts fail.

The fast path begins by using the point query \texttt{sl.ceil(l)} to find the first logically present node at or after the start of the range (line \ref{lst-fast-path-ceil}).\footnote{We say ``at or after the start of the range'' rather than ``within the range'' because it is possible that there are no nodes within the range.}
It then scans the skip list, copying the keys and values of all logically present nodes along the way (line \ref{lst-fast-add}), until it reaches a node beyond the end of the range (possibly \texttt{sl.tail}).

\subsubsection{Slow Path}

The slow path is more nuanced. 
First, a transaction is used to initialize the range query (line \ref{lst-slow-path}). 
In addition to finding the start node using \texttt{sl.ceil(l)}, it also calls \texttt{rqc.on\_range()}.
This informs the RQC of its existence and gets a version number, $\texttt{ver}$. 
Doing both tasks in one transaction ensures that \texttt{start} is a safe node for the query.

After this setup is complete, \texttt{range\_slow()} (line \ref{lst-range-slow-method}) is responsible for collecting key/value pairs.
The variables \texttt{set} and \texttt{n} record the progress of the range query---the set of collected pairs and the current node, respectively. 
We use the syntax \texttt{atomic}\-\texttt{(no\_local\_undo)} to indicate that the STM should not roll back modifications to these local variables upon transaction abort (line \ref{lst-no-local-undo}).
Due to the fact that \texttt{range\_slow()} does not roll back these local variables or modify any shared variables, it is able to keep all progress up to the point where the transaction aborted---thus turning aborts into unexpected early commits. 
Furthermore, \texttt{n} is only ever set to a safe node. 
Therefore, the next transaction is always able to pick up right where the last one left off. 
The somewhat verbose syntax on lines~\ref{lst-range-process-loop-start}--\ref{lst-range-process-loop-end} employs the clearly defined abort points of STM to ensure no key/value pair is inserted into the set twice.

\texttt{next\_safe()} (line \ref{lst-next-safe-method}) traverses the bottom level of the skip list until it finds the next safe node after \texttt{n}.
Since \texttt{sl.tail} is always safe, and the method is never called with \texttt{sl.tail} as its argument, such a node always exists.
\texttt{is\_safe()} (line \ref{lst-is-safe-method}) determines if a node is safe for a range query with version number \texttt{ver}, using the criteria from Section~\ref{subsec-safe-nodes}.

On line \ref{lst-slow-after-range} the slow-path range query informs the RQC that it is complete by calling \texttt{rqc.after}\-\texttt{\_range()}. 
This tells the RQC that it no longer needs its safe nodes, which may trigger the unstitching and reclamation of nodes whose removal was previously deferred.

\subsection{Implementing the Range Query Coordinator}

\begin{figure}
\vspace{-4mm}
\begin{lstlisting}[escapechar=`, firstnumber=1]
type rqc # The RQC implementation `\label{lst-rqc-def}`
  counter: u64 # Counter for generating version numbers
  range_ops: LinkedList<range_op>

type range_op # Metadata about a slow-path range op `\label{lst-range-op-def}`
  const ver: u64 # Version number of the range op
  deferred: LinkedList<sl_node> # Nodes to remove

# Inform the RQC about a slow-path range query
func on_range() -> u64:
  range_ops.append(new range_op(++counter, [])) `\label{lst-on-range-magic}`
  return counter

# Get a version number for insertion/removal time
func on_update() -> u64: 
  return counter `\label{lst-on-update-body}`

# Give a node to the RQC for removal when safe
func after_remove(n: sl_node): `\label{lst-after-remove-method}`
  atomic:
    let tail = range_ops.tail() # Returns null if empty
    if !tail || n.i_time >= tail.ver: `\label{lst-after-remove-immediate-check}`
      n.unstitch() # Safe to remove immediately
      delete n
    else:
      tail.deferred.append(n) # Defer removal `\label{lst-rqc-impl-defer-removal}`
            
# Clean up after a slow-path range op
func after_range(ver: u64): `\label{lst-after-range-method}`
  let removals = [] # Nodes to remove immediately
  atomic:
    let op = range_ops.find(ver) # Find by version num.
    let pred = range_ops.pred(op) # Point query
    range_ops.remove(op) `\label{lst-after-range-remove-op}`
    if (!pred): # Take responsibility for removing now
      removals = op.deferred `\label{lst-take-deferrals}`
    else: # Defer removals further 
      pred.deferred.append_all(op.deferred) `\label{lst-defer-further}`
    delete op
  for n in removals: # Remove the deferred nodes  `\label{lst-clean-deferrals-start}`
    atomic:
      n.unstitch() `\label{lst-clean-deferrals-end}`
\end{lstlisting}
\vspace{-5mm}
\caption{A concrete implementation of the range query coordinator.}
\label{lst-clock-impl}
\end{figure}

Figure~\ref{lst-clock-impl} presents a concrete implementation of the RQC\@.
The RQC itself (line \ref{lst-rqc-def}) has two fields: \texttt{counter}, a 64-bit integer for generating version numbers; and \texttt{range\_ops}, a doubly linked list of all ongoing slow-path range queries. 
The list contains objects of type \texttt{range\_op} (line \ref{lst-range-op-def}), which consist of two fields: a version number (\texttt{ver}) and a set of logically deleted nodes whose removal has been deferred until after the range query is complete (\texttt{deferred}).

The simplest algorithm would assign a unique version number to each operation, but \texttt{counter} risks becoming a contention hotspot, with every write to it potentially causing aborts for concurrent transactions that interacted with it. 
Instead, we only increment the counter for range queries (\texttt{on\_range()}, line \ref{lst-on-range-magic}).
Elemental operations reuse the most recent range query's version number (\texttt{on\_update()}, line \ref{lst-on-update-body}), thus ordering themselves after it.
\texttt{on\_range()} also creates a new \texttt{range\_op} object and adds it to the end of the list. 

While consecutive range queries could also share version numbers, we did not see any benefit to doing so: 
Short range queries should complete on the fast path without accessing the RQC, and long range queries should perform enough real work that their RQC increments should not contend.

In \texttt{after\_remove()} (line \ref{lst-after-remove-method}), the RQC assumes responsibility for unstitching and deleting a logically removed node $n$. 
First, it checks if the removal can be done immediately (line \ref{lst-after-remove-immediate-check}).
It can do so if either \texttt{range\_ops} is empty or $n$ was inserted after the most recent ongoing slow-path range query $R_{last}$. 
Otherwise, $n$ is a safe node for $R_{last}$, so $n$ is added to its \texttt{deferred} list (line \ref{lst-rqc-impl-defer-removal}).

The most complex logic is in \texttt{after\_range()} (line \ref{lst-after-range-method}). 
First, the \texttt{range\_op} representing the finishing range query is removed from the list (line \ref{lst-after-range-remove-op}). 
Then, deferred removals must be handled.
There are two scenarios here.
(1) If the removed \texttt{range\_op} is the oldest one (the head of \texttt{range\_ops}), then the nodes in \texttt{deferred} can now be safely unstitched and reclaimed (lines~\ref{lst-take-deferrals}, \ref{lst-clean-deferrals-start}--\ref{lst-clean-deferrals-end}). 
(2) However, if a \texttt{range\_op} with an earlier version number still exists, one of its safe nodes may be in \texttt{deferred}.
The potential benefit of scanning the entries in \texttt{deferred} and comparing their removal times to the remaining \texttt{range\_ops}' versions is not worth the overhead.
Instead, the removal of all of these nodes is deferred further by adding them to the predecessor's \texttt{deferred} list, via an $O(1)$ list append operation. (line \ref{lst-defer-further})
Since these nodes are being passed backward, not forward, every node is guaranteed to be reclaimed eventually.

When many removals run concurrently with a slow-path range query $R$, insertions into $R$'s \texttt{deferred} list can become a contention bottleneck.
By virtue of the above mechanism, it is safe to delegate unstitching and reclaiming a range query's safe node to a later range query.
We leverage this fact by keeping a buffer of removed nodes for each thread.
We can then modify line \ref{lst-rqc-impl-defer-removal} in \texttt{after\_range()} to instead push the node into that thread's buffer.
When the buffer is full (size $32$ in our implementation), the thread checks if there are any active slow-path range queries.
If not, it can immediately unstitch and reclaim all entries.
Otherwise, it transfers the entire buffer to the most recent range query's \texttt{deferred} list via an $O(1)$ append operation.

\subsection{Correctness}
Here we briefly outline the properties that would underpin a correctness proof.
First, we observe that each elemental and point query operation executes as a single transaction.
Thus as long as each would be correct in a sequential implementation, they are all correct in a concurrent execution.
Next, we note that a fast-path range query works correctly because it is also a single transaction.
The node found by \texttt{ceil} will not be unstitched, nor can nodes within the range be added or removed, without causing the entire range query to abort.

The correctness of a slow-path range query $R$ is more complex.
The first atomic step uses \texttt{ceil} to find a logically present node and then increments the version number.
This ensures that the node will not be unstitched before $R$ completes.
This is also $R$'s linearization point.
Then, $R$ only pauses at nodes that cannot be unstitched until after it completes.
The unstitching of an unsafe node while $R$ is accessing it would cause a transaction abort, causing $R$ to retry from the last safe node.
Since all the unstitched nodes have a removal time less than the version number of our range query, the query will not miss any necessary nodes.
Finally, the \texttt{i\_time} field ensures that the range query does not include nodes inserted after its linearization point.

\section{Evaluation}
\label{sec-eval}

In this section, we compare the performance of the skip hash against a set of ordered map implementations that represent the current state of the art.
These include a binary search tree and skip list based on the versioned compare-and-swap (vCAS) technique~\cite{wei2021constant} and a skip list that uses bundled references~\cite{nelson2022bundling}.
We also considered variants of these three maps that utilize a hardware timestamp counter instead of a shared memory counter, using the x86 \texttt{rdtscp} instruction~\cite{grimes2023rdtscp}.
This eliminates a known contention bottleneck for vCAS and bundling.
This optimization improves performance in all evaluated workloads, and so we have excluded the non-optimized variants of these maps from all charts.
Lastly, in workloads consisting entirely of elemental operations, we evaluate a hashmap and doubly linked skip list that do not support range queries, implemented using STM.
\begin{figure*}
    \centering
    \begin{subfigure}[b]{0.24\textwidth}
        \centering
        \includegraphics[width=\textwidth]{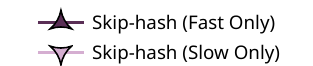}
    \end{subfigure}
    \begin{subfigure}[b]{0.24\textwidth}
        \centering
        \includegraphics[width=\textwidth]{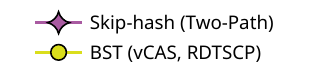}
    \end{subfigure}
    \begin{subfigure}[b]{0.24\textwidth}
        \centering
        \includegraphics[width=\textwidth]{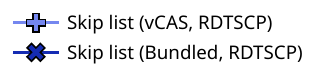}
    \end{subfigure}
    \begin{subfigure}[b]{0.24\textwidth}
        \centering
        \includegraphics[width=\textwidth]{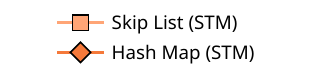}
    \end{subfigure}
    \hfill
    \begin{subfigure}[b]{0.33\textwidth}
        \centering
        \includegraphics[width=\textwidth]{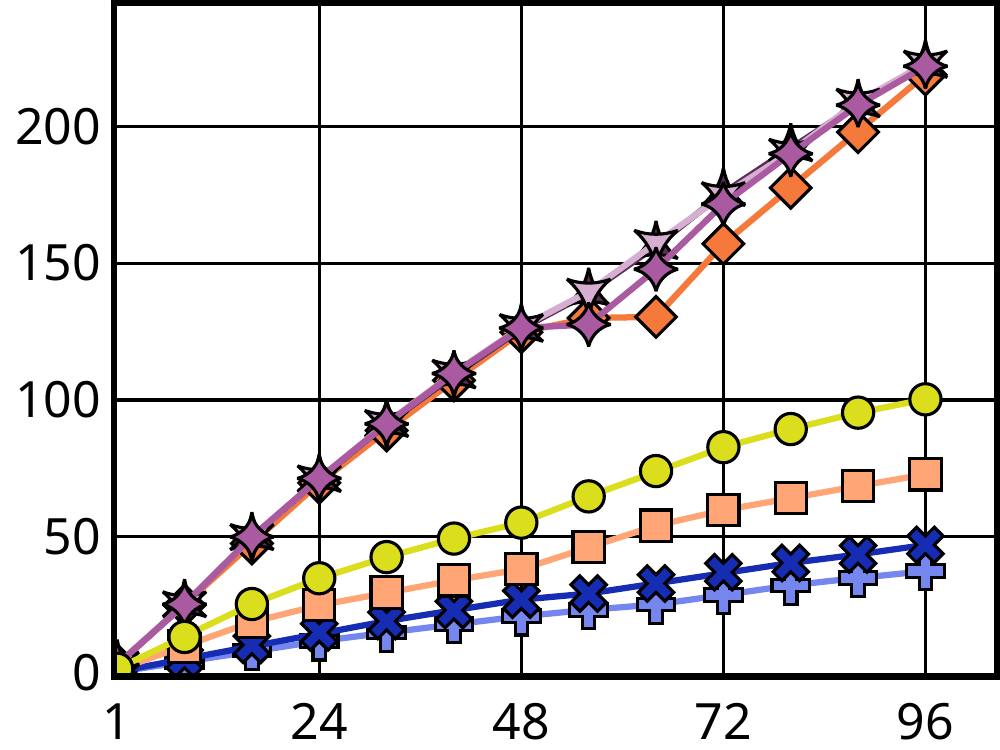}
        \caption{100\% lookup}
        \label{chart-100-0-0}
    \end{subfigure}
    \begin{subfigure}[b]{0.33\textwidth}
        \centering
        \includegraphics[width=\textwidth]{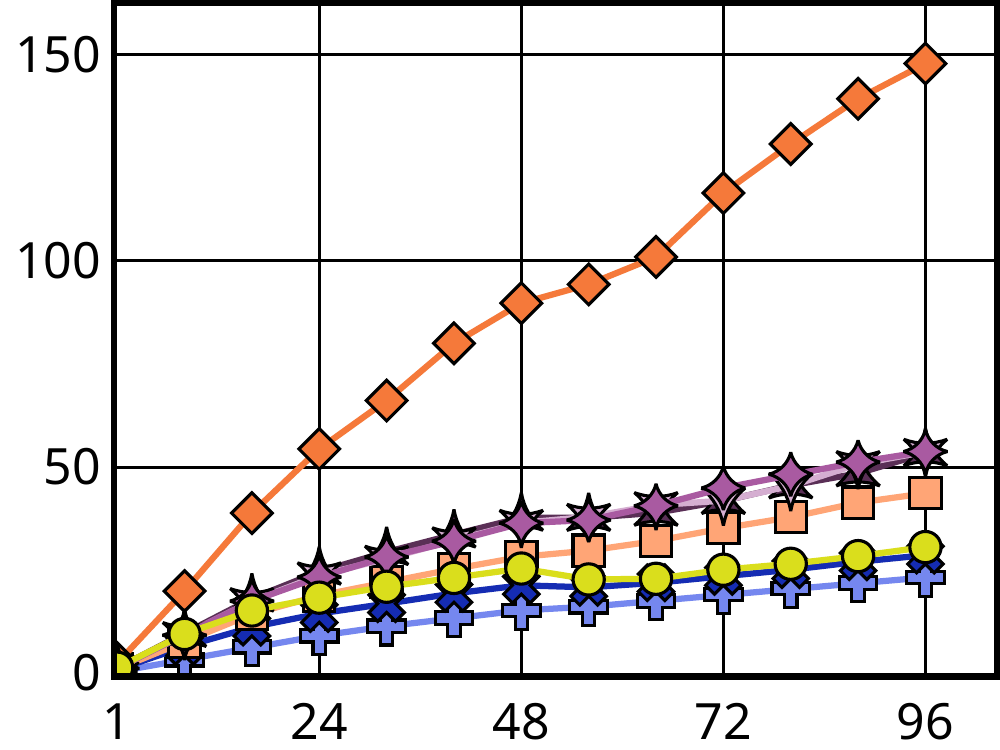}
        \caption{100\% update}
        \label{chart-0-100-0}
    \end{subfigure}
    \begin{subfigure}[b]{0.33\textwidth}
        \centering
        \includegraphics[width=\textwidth]{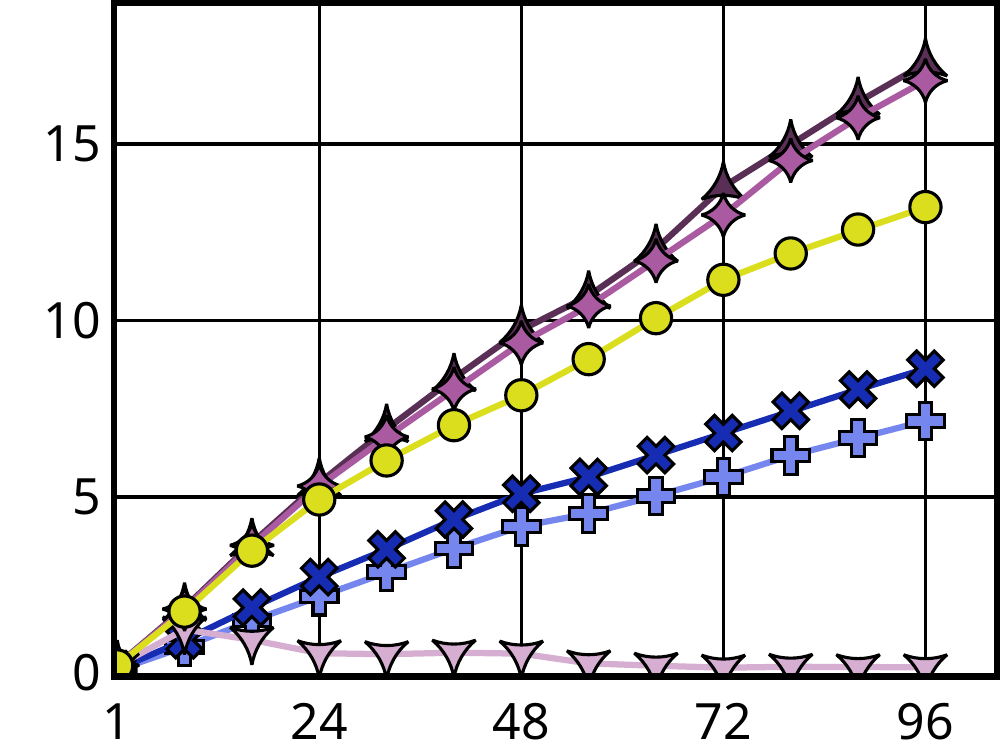}
        \caption{100\% range}
        \label{chart-0-0-100-100}
    \end{subfigure}
    \hfill
    \begin{subfigure}[b]{0.33\textwidth}
        \centering
        \includegraphics[width=\textwidth]{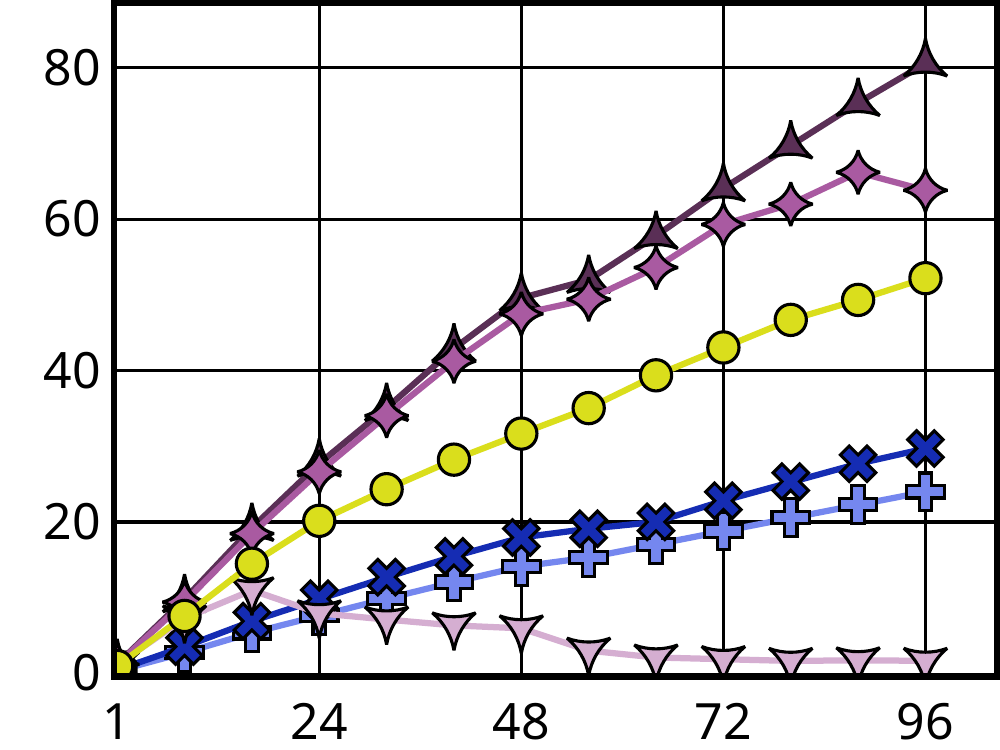}
        \caption{80\% lookup, 10\% update, 10\% range}
        \label{chart-80-10-10-100}
    \end{subfigure}
    \begin{subfigure}[b]{0.33\textwidth}
        \centering
        \includegraphics[width=\textwidth]{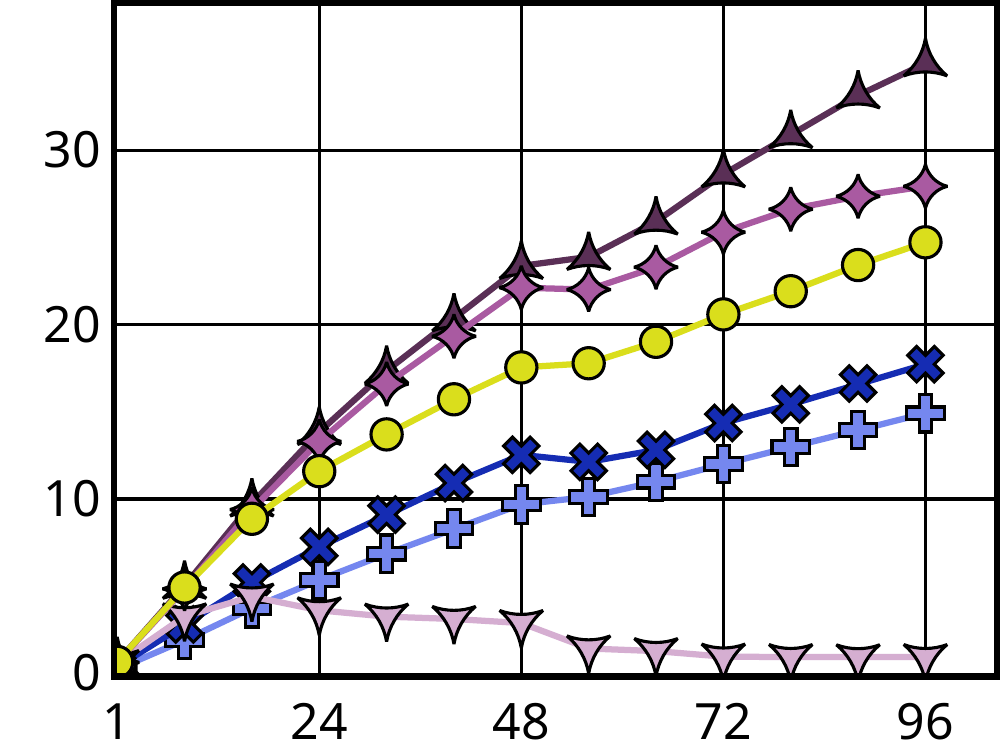}
        \caption{80\% update, 20\% range}
        \label{chart-0-80-20-100}
    \end{subfigure}
    \begin{subfigure}[b]{0.33\textwidth}
        \centering
        \includegraphics[width=\textwidth]{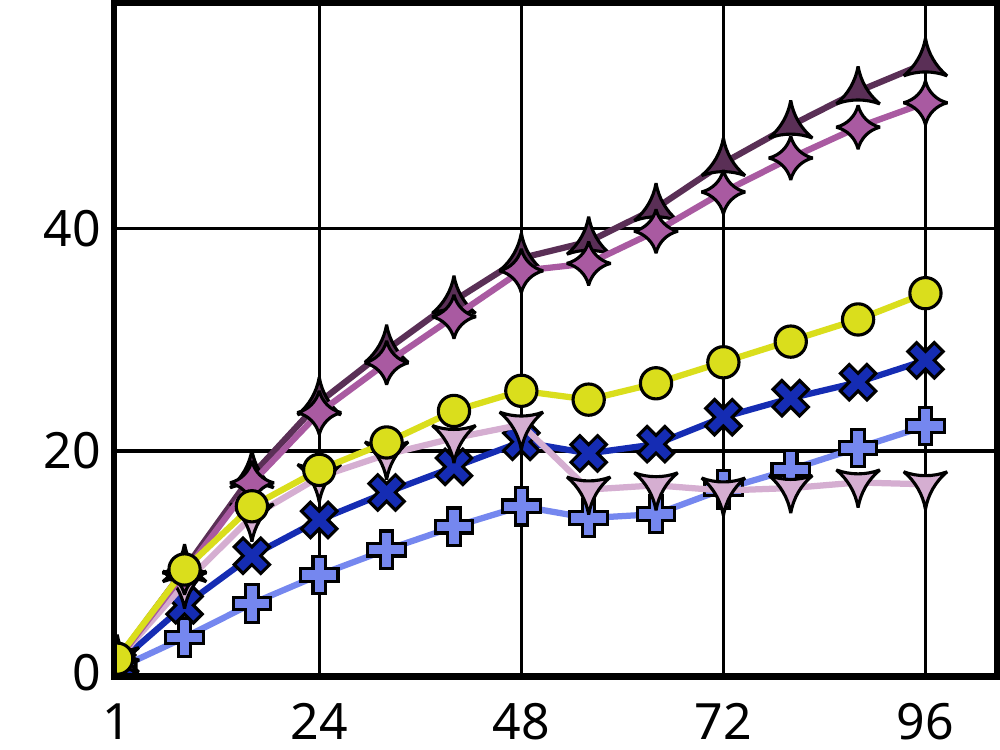}
        \caption{1\% lookup, 98\% update, 1\% range}
        \label{chart-1-98-1-100}
    \end{subfigure}
    \caption{Comparison of skip hash performance versus state-of-the-art ordered maps, with varying mixtures of lookup, update, and range operations. The x-axis represents thread count, and the y-axis represents throughput as millions of operations per second. All range queries were of length $100$, thus processing $50$ keys on average.}
    \label{eval-charts-workloads}
\end{figure*}

\subsection{Experimental Setup}
All experiments were conducted on a system running Ubuntu 22.04 with 180 GB of RAM and two Intel Xeon Platinum 8160 CPUs running at 2.10 GHz. 
Each CPU has 24 cores, for a total of 48 cores and 96 hardware threads. 
For all maps evaluated, threads were pinned to cores in the same way. 
The first 24 threads were pinned to unique cores on a single CPU; the next 24 threads were pinned to the second thread on those cores; and the last 48 threads were distributed in the same manner on the second CPU. 
Thus, in all charts, symmetric multithreading is not a factor for data points up through 24 threads, and non-uniform memory access (NUMA) latency is not a factor up through 48 threads.

We used the evaluation framework from~\cite{grimes2023rdtscp}, which already included support for the vCAS and bundling-based data structures, as well as their \texttt{rdtscp}-enhanced variants.
The default behavior of this framework is that worker threads perform lookups, insertions, removals, and range queries in proportions that vary by workload. 
Keys and values were both represented as signed 64-bit integers.
In this paper, we report results for a key universe of $10^6$.
In experiments with larger universes, the trends are the same, but the separation between the skip hash and other data structures is more pronounced, particularly for elemental operations.

Before each experiment, each map was pre-filled with half of the keys in the universe, resulting in a population of $5 \cdot 10^5$. 
Elemental operations chose keys uniformly at random from the universe. 
Range queries did the same to choose $l$ and computed $r$ by adding a fixed range length determined by the particular workload. 
They then copied all keys and values within that range into a pre-allocated buffer. 
In all workloads, update operations were evenly split between insertions and removals, so that the population of the data structure remained roughly constant at all times with high probability. 
This also meant that all elemental operations were equally likely to succeed as fail, and each range query was expected to process $(r-l)/2$ entries. 
Unless stated otherwise, the range length was $100$ keys (as in~\cite{nelson2022bundling, grimes2023rdtscp}), for an expected $50$ entries processed.
All benchmark and map code was written in C++ and compiled with g++ version 12.3.0 using flags \texttt{-O3 -std=c++20}.
All experiments used the jemalloc~\cite{jemalloc-website-2017} allocator.
All data points are the average of five 3-second trials.
Error bars are omitted, because we did not observe significant variance.

In the experiments, we considered three implementations of the skip hash: one where all range queries use the fast path, one where they all use the slow path, and one where they fall back to the slow path after three fast-path failures.
Following common guidance that hash tables perform best when about $70\%$ full and that prime hash table sizes are advantageous~\cite{clrs-book-3ed}, we configured each skip hash's hash table to consist of $714,341$ buckets. 
This number was chosen due to being the smallest prime that yields a hash table utilization rate $\leq 70\%$ for the expected population.
Keys were hashed using \texttt{std::hash}.
For the skip hash and skip lists, the level count was set to $20$, as $2^{20}$ is slightly greater than $10^6$.

The STM implementation used in the skip hash, skip list, and hash map is exoTM~\cite{sheng-pact-2023}.
Since conflicts are expected to be rare, we chose the eager (undo) algorithm \emph{without} timestamp extension.
This is expected to have the lowest latency, though it has the highest abort overhead of any STM algorithm in the exoTM framework.
In the charts, we present results using an \texttt{rdtscp}-based clock~\cite{ruan-taco-2013}.
We also tested the gv1 and gv5 logical clocks~\cite{dice-disc-2006, lev-spaa-2009}.
Gv1 did not scale well for the skip hash's small transactions, while gv5 and the slow clock performed poorly for slow-path experiments, where our RQC implementation violates the assumptions that underpin their designs.

\subsection{Analysis}
Figure~\ref{eval-charts-workloads} presents throughput as the thread count is varied from 1 up to the maximum hardware thread count in a microbenchmark with various workloads.
We begin by discussing workloads that perform a single operation in isolation to evaluate the raw performance of those operations. 

\subsubsection{Isolated Workloads}
\label{sec-eval-isolated}
In a workload consisting of 100\% lookup operations (Figure~\ref{chart-100-0-0}), the skip hash scales to the maximum thread count and achieves a $>2\times$ speedup over all data structures except the hash map, which does not support range queries.
This is because the \texttt{lookup()} operation is $O(1)$ for the skip hash and hash map, as opposed to $O(\log{n})$ for skip lists and BSTs. 
Furthermore, our STM implementation does not suffer aborts in such read-only workloads.
We also observe that the STM skip list outperforms skip lists with more complex synchronization mechanisms, which confirms our claim that modern STM can be fast and shows the benefit of composing a hash map and skip list.

Next, in Figure~\ref{chart-0-100-0}, we evaluate the performance of update operations (an equal mix of insertions and removals). 
The throughput of the skip hash drops due to the additional overhead of stitching and unstitching (and rare transaction aborts). 
However, it maintains a significant lead over all data structures save the hash map, as hash acceleration allows the skip hash to avoid $O(\log{n})$ skip list searches for all operations except successful insertions. 
Furthermore, we see that the slow path mechanism does not introduce significant overhead when it is not in use.
Note that in these charts and many that follow, a small performance dip occurs for some maps after 48 threads, due to the cross-chip memory traffic.

Lastly, we evaluate the performance of range queries of length $100$ in Figure~\ref{chart-0-0-100-100}. 
Note that in this figure, the y-axis represents completed range queries, not the total number of nodes accessed.
As this is a read-only workload, no transactions will abort, so the two-path variant will complete all range queries on the fast path, thus avoiding any contention on the RQC\@.
Therefore, the fast-only and two-path skip hashes are able to outperform all other data structures by a significant margin.
However, the slow-only skip hash suffers from contention on the RQC, leading to performance degradation.
Allowing range queries to reuse version numbers would not help much, as they would still contend when inserting into \texttt{range\_ops}.
This behavior is a consequence of the small range size.
We study it further in Section~\ref{sec-eval-size}.

\subsubsection{Mixed Workloads}
\label{sec-eval-mixed}

The main difficulty with range queries is not running them in isolation---which can be done without any special mechanism---but keeping them both linearizable and performant in the face of concurrent updates. 
Therefore, in Figures~\ref{chart-80-10-10-100}--\ref{chart-1-98-1-100}, we evaluate workloads that do both.
As before, range queries access $50$ elements on average.

For each workload in this section, worker threads choose what type of operation to do at random, based on a distribution specified by that workload. 
For this reason, it does not make sense to separate throughput by operation type.

First, Figure~\ref{chart-80-10-10-100} shows throughput as thread count is varied for a workload with a 10\% updates and 10\% range queries.
The fast-only and two-path skip hashes maintain a moderate lead over the vCAS BST and a $>2\times$ speedup over all other data structures.
The difference in performance results from a small fraction of range queries needing more than $3$ attempts to complete as a single STM transaction.
In the fast-only algorithm, these additional attempts are made and succeed.
In the two-path algorithm, the slow path is chosen instead, resulting in higher latency.

In Figure~\ref{chart-0-80-20-100}, each thread performs 80\% update operations and 20\% range queries. 
Since each range query counts as $1$ operation, all numbers are significantly lower than in the previous chart.
However, the general trends are otherwise the same: Increasing the update ratio does not impact fast-path range queries, but slow-path range queries continue to suffer.
We profiled this overhead, and found that it was a result of two factors.
First, slow-path range queries were competing to attain version numbers.
Second, removal operations were aborting due to contention as they tried to delegate unstitching to in-flight range queries.

Lastly, in Figure~\ref{chart-1-98-1-100}, we consider a workload with a much greater proportion of update operations per range query (98 instead of 4).
This change greatly increases the probability a given range query will be aborted by a concurrent update but reduces contention on global metadata on the slow path.
The slow-path throughput improves significantly but not enough to be competitive.
Apart from that, all trends are the same.
We found that remove operations are not as common a source of aborts, but the slow-path range queries were still competing to acquire version numbers.
In short, we conclude that our RQC mechanism is a bottleneck for short-running range queries due to interactions with concurrent range queries, not concurrent removals.
However, the two-path approach all but eliminates the need for a slow path for this kind of workload.

\subsubsection{Varying Range Query Size}
\label{sec-eval-size}

From the above discussion, it may seem that there is no utility in having a slow path.
Recall, however, that starvation is a well-known problem for long-running read-only transactions~\cite{bobba-isca-2007}, and in the previous section, range queries were short. 
To understand how range query length affects performance of the evaluated maps, we conducted an experiment in which we varied that parameter rather than the thread count. 
Thread count was held constant at 48, split evenly into 24 update-only threads and 24 range-only threads, with all threads on one socket.

\begin{figure}[b]
    \centering
    \includegraphics[width=\columnwidth]{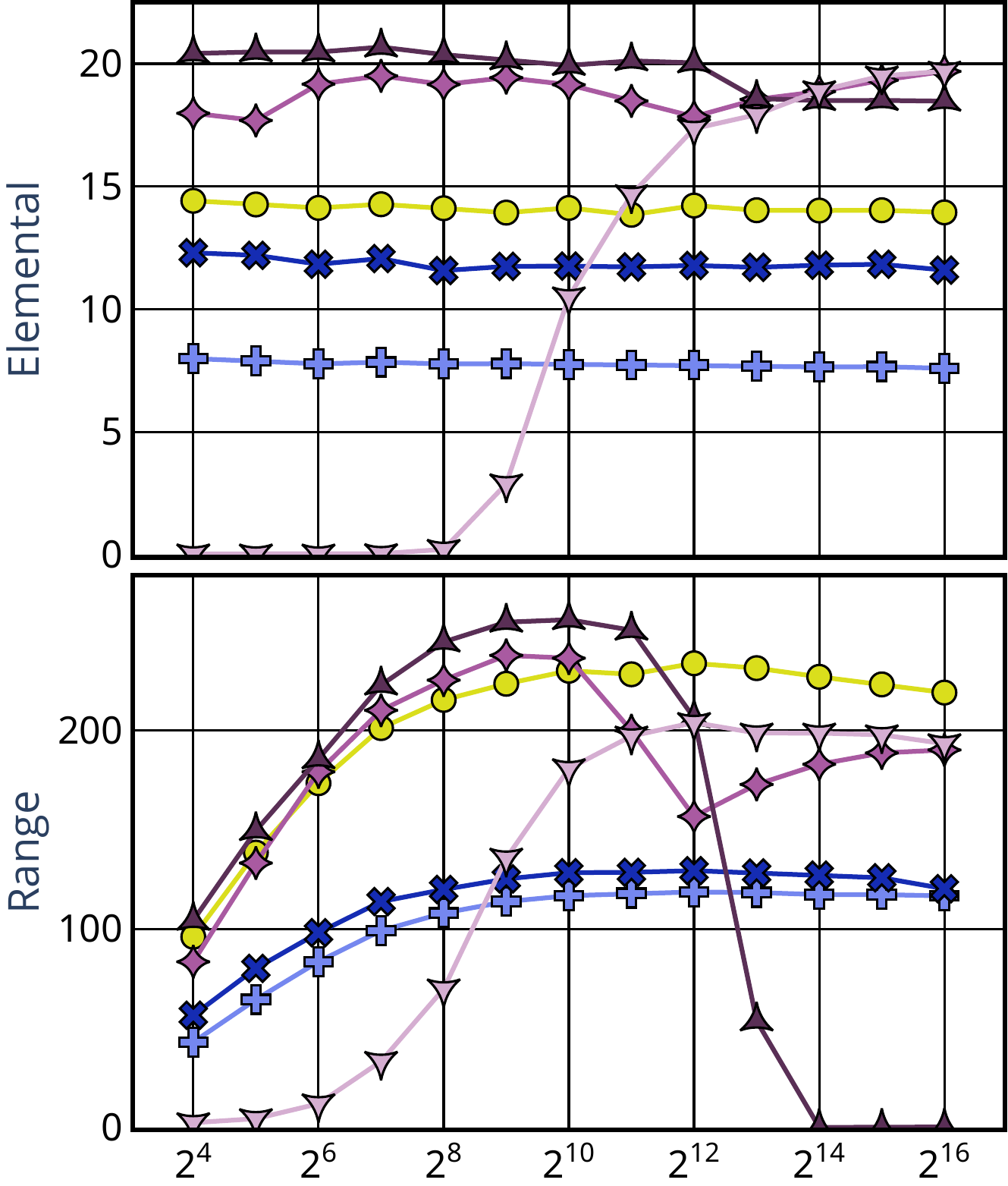}
    \caption{Experiments with 24 update-only and 24 range-only threads as range query length varies. Top: update throughput (in millions of ops/sec). Bottom: range query throughput (in millions of nodes processed/sec).}
    \label{chart-range-lengths}
\end{figure}

Figure~\ref{chart-range-lengths} shows the results of this experiment. 
Since different operation types are performed by different threads, their throughput can vary independently, so the figure consists of two charts. 
The upper chart depicts update throughput in millions of operations per second, as before.
The lower chart measures the performance of range queries but, instead of counting the number of range queries performed, it shows \emph{how many key/value pairs were processed}.
This improves readability at large range lengths.

The throughput of update operations remains almost constant across range lengths for most maps. 
For the skip hashes with fast-only and two-path range queries, their performance remains consistent and significantly faster than all data structures.
The only surprise in this figure is for the slow-only skip hash.
Here, we observe that elemental throughput improves for ranges of length $\geq 2^9$, eventually matching the peak throughput achieved by the other skip hashes.

The chart showing range throughput, on the other hand, is not flat. 
All evaluated maps improve in performance from $2^4$ to $2^{10}$, because they amortize the boundary overheads of a range query over a greater number of elements accessed.
That is, all evaluated maps have some overhead involved in setting up and/or tearing down a range query. 
As range queries become longer, they are able to spend a proportionately greater time processing nodes. 
The performance of prior work plateaus once the length of the range queries becomes large enough to make this overhead irrelevant.
However, the performance of the fast-path skip hash degrades rapidly for range queries starting at $2^{11}$.
This is after the point at which slow-path range query performance improved, and so the two-path skip hash is able to use the slow-path after three failed fast-path attempts.

\begin{table}
    \begin{tabular}{r||r|r|r|r|r||}
    Range Length & $2^{10}$& $2^{11}$& $2^{12}$& $2^{13}$ & $2^{14}$\\ \hline
    Abort Rate & 1.07 & 1.34 & 2.66 & 22.8 & $\infty$\\
    \end{tabular}
    \caption{Aborts per successful range query in a fast-path-only skip hash, by range length.}
    \label{eval-table-path-frequency}
\end{table}

Table~\ref{eval-table-path-frequency} reports the average number of aborts per successful range query as a function of the range query size in the fast-only variant.
We see a precipitous increase in aborts as the query size increases until, at a range size of $2^{14}$, no range query is able to complete.
These aborts require some explanation.
Recall that read-only STM transactions only abort if they (1) encounter an element that was updated after they began, and (2) they cannot prove that everything they read before that encounter is unchanged.
In our highly optimized implementation of read-only transactions, there is no support for disproving condition (2).
So then, when a fast-path range transaction encounters a modification that occurred after it began, it cannot prove that it can linearize, so it aborts.
Were we to enable condition (2), latency would increase for all range queries, but in a uniform workload the expected impact would only be a factor of two improvement in range size before the same abort frequency occurs.
In short, long-running fast-path range queries do not have an easy way to avoid starvation.
Fortunately, even our crude mechanism for transitioning to the slow path alleviates most of this overhead.
Note that this experiment motivates transitioning to the slow path earlier, while previous experiments favored moving to the slow path later.
Should practitioners choose to use the skip hash, we recommend they explore more nuanced strategies for switching to the slow path.

\section{Conclusions and Future Work}
\label{sec-conclusion}

In this paper, we introduced the skip hash, which uses modern software transactional memory (STM) to compose a closed-addressing hash map with a doubly linked skip list.
The resulting data structure has $O(1)$ complexity for most operations, resulting in exceptional throughput for elemental and point-query methods.
By virtue of its use of STM, the skip hash easily supports linearizable range queries, but these risk starvation for high-contention workloads.
To overcome this problem, we introduced a lightweight versioning technique that allows linearizable range queries to ignore values added after they linearized and that also defers reclamation of logically deleted nodes that they may still need.

In experimental evaluation, we saw that the skip hash outperformed the state of the art in almost every workload configuration, and often by a large margin.
We also showed that our fast-path/slow-path mechanism is effective in preventing starvation for long-running range queries.

The skip hash design is unconventional, both in its use of double-linking and in the way it composes data structures.
However, its implementation is not complicated: the entire data structure requires under 1000 lines of code.
These properties are a direct consequence of our decision to treat fast STM as a given, and assume that atomic multi-word transactions would ``just work''.
In short, using STM simplified our design, implementation, and verification tasks.

There are several exciting directions for future research.
First and foremost, we plan to study whether a hardware clock could avoid contention on RQC metadata.
We hypothesize that integrating techniques developed by Grimes et al.~\cite{grimes2023rdtscp} into the RQC could eliminate the contention we observed for short slow-path range queries.
As a second direction, since our work has shown that low-level STM implementation details do matter, there may be yet-undiscovered optimizations specific to STM-based data structure design.
Lastly, the skip hash's asymptotic complexity makes it an appealing candidate for distributed and persistent memory systems.
These systems have higher average memory access latencies than traditional shared memory multiprocessors, so reducing the number of memory accesses is essential to achieving good performance.
This will, of course, necessitate a renewed focus on distributed and/or persistent STM, particularly one that incorporates assumptions analogous to those that resulted in the fast STM systems that facilitated this research.
\clearpage
\bibliography{mfs}
\end{document}